\documentclass[sigconf]{acmart}

\AtBeginDocument{%
 }

\setcopyright{acmlicensed}
\copyrightyear{2018}
\acmYear{2018}
\acmDOI{XXXXXXX.XXXXXXX}

\acmConference[Conference acronym 'XX]{Make sure to enter the correct
  conference title from your rights confirmation emai}{June 03--05,
  2018}{Woodstock, NY}
\acmISBN{978-1-4503-XXXX-X/18/06}

\usepackage{multirow}
\usepackage{graphicx}
\usepackage{subcaption}
\usepackage{booktabs}
\usepackage{arydshln}
\usepackage{xcolor}
\usepackage{amsmath}
\setcopyright{none}
\renewcommand\footnotetextcopyrightpermission[1]{}

\begin{document}

\title{Stop Treating Collisions Equally: Qualification-Aware Semantic ID Learning for Recommendation at Industrial Scale}


\author{Zheng Hu}
\authornotemark[2]
\affiliation{%
  \institution{University of Electronic Science and Technology of China}
  \city{Chengdu}
  \country{China}
}

\author{Yuxin Chen}
\authornote{Corresponding authors.}
\affiliation{%
  \institution{Kuaishou Technology}
  \city{Beijing}
  \country{China}
}
\email{chenyuxin06@kuaishou.com}

\author{Yongsen Pan}
\authornote{Work done during internship at Kuaishou Technology.}
\affiliation{%
  \institution{University of Electronic Science and Technology of China}
  \city{Chengdu}
  \country{China}
}

\author{Xu Yuan}
\author{Yuting Yin}
\author{Daoyuan Wang}
\affiliation{%
  \institution{Kuaishou Technology}
  \city{Beijing}
  \country{China}
}

\author{Boyang Xia}
\author{Zefei Luo}
\affiliation{%
  \institution{Kuaishou Technology}
  \city{Beijing}
  \country{China}
}

\author{Hongyang Wang}
\author{Songhao Ni}
\affiliation{%
  \institution{Kuaishou Technology}
  \city{Beijing}
  \country{China}
}

\author{Dongxu Liang}
\affiliation{%
  \institution{Kuaishou Technology}
  \city{Beijing}
  \country{China}
}

\author{Jun Wang}
\affiliation{%
  \institution{Kuaishou Technology}
  \city{Beijing}
  \country{China}
}

\author{Shimin Cai}
\affiliation{%
  \institution{University of Electronic Science and Technology of China}
  \city{Chengdu}
  \country{China}
}

\author{Tao Zhou}
\affiliation{%
  \institution{University of Electronic Science and Technology of China}
  \city{Chengdu}
  \country{China}
}

\author{Fuji Ren}
\authornotemark[1] 
\affiliation{%
  \institution{University of Electronic Science and Technology of China}
  \city{Chengdu}
  \country{China}
}
\email{renfuji@uestc.edu.cn}

\author{Wenwu Ou}
\affiliation{%
  \institution{Kuaishou Technology}
  \city{Beijing}
  \country{China}
}

\begin{abstract}
Semantic IDs (SIDs) are compact discrete representations derived from multimodal item features, serving as a unified abstraction for ID-based and generative recommendation. However, learning high-quality SIDs remains challenging due to two issues. \textbf{(1) Collision problem}: the quantized token space is prone to collisions, in which semantically distinct items are assigned identical or overly similar SID compositions, resulting in semantic entanglement. \textbf{(2) Collision-signal heterogeneity}: collisions are not uniformly harmful. Some reflect genuine conflicts between semantically unrelated items, while others stem from benign redundancy or systematic data effects. To address these challenges, we propose \textbf{Qualification-Aware Semantic ID Learning (QuaSID)}, an end-to-end framework that learns collision-qualified SIDs by selectively repelling qualified conflict pairs and scaling the repulsion strength by collision severity. QuaSID consists of two mechanisms: Hamming-guided Margin Repulsion, which translates low-Hamming SID overlaps into explicit, severity-scaled geometric constraints on the encoder space; and Conflict-Aware Valid Pair Masking, which masks protocol-induced benign overlaps to denoise repulsion supervision. In addition, QuaSID incorporates a dual-tower contrastive objective to inject collaborative signals into tokenization. Experiments on public benchmarks and industrial data validate QuaSID. On public datasets, QuaSID consistently outperforms strong baselines, improving top-K ranking quality by \textbf{5.9\%} over the best baseline while increasing SID composition diversity. In an online A/B test on Kuaishou e-commerce with a 5\% traffic split, QuaSID increases ranking GMV-S2 by \textbf{2.38\%} and improves completed orders on cold-start retrieval by up to \textbf{6.42\%}. Finally, we show that the proposed repulsion loss is plug-and-play and enhances a range of SID learning frameworks across datasets.

\end{abstract}



\maketitle
\section{Introduction}
In modern recommendation systems, Semantic IDs (SIDs) have emerged as a promising mechanism for representing items using compact, discrete codes distilled from rich multimodal features such as text, images, audio, and structured metadata \cite{DBLP:conf/kdd/Hou0SJSSHYM25,DBLP:conf/kdd/00050ZW25,DBLP:conf/kdd/BinCYZHY0ZYL25,DBLP:conf/icde/ZhengHLCZCW24}. SIDs can enhance both traditional recommendation pipelines and emerging generative recommendation systems \cite{DBLP:journals/corr/abs-2502-12448, li2025survey}. Unlike conventional ID embeddings that depend on ever-growing embedding tables and ad-hoc hashing, SID-based recommendation represents each item as a fixed-length sequence of discrete semantic tokens \cite{DBLP:conf/recsys/SinghV0KSZHHWTC24}. This yields compact, reusable, and index-friendly identifiers that mitigate vocabulary explosion in both retrieval and generation, while reducing hash-induced collision noise and stabilizing item identities under frequent item updates, re-indexing, and real-world life-cycle drift \cite{DBLP:conf/recsys/ZhengHPRWXNL00L25}. These properties make SIDs particularly valuable in large-scale industrial environments, where efficiency, interpretability, and cross-system consistency are essential. Moreover, as the field shifts from classical retrieval-and-ranking architectures toward generative recommendation systems (GRS), the importance of SIDs continues to grow \cite{DBLP:journals/corr/abs-2506-13695,DBLP:journals/corr/abs-2508-20900}. By providing a unified tokenized interface, SIDs allow items to be seamlessly incorporated into generative models, enabling them to generate recommendations in the same manner as they generate language.

\begin{figure}
    \centering
    \includegraphics[width=1.0\linewidth]{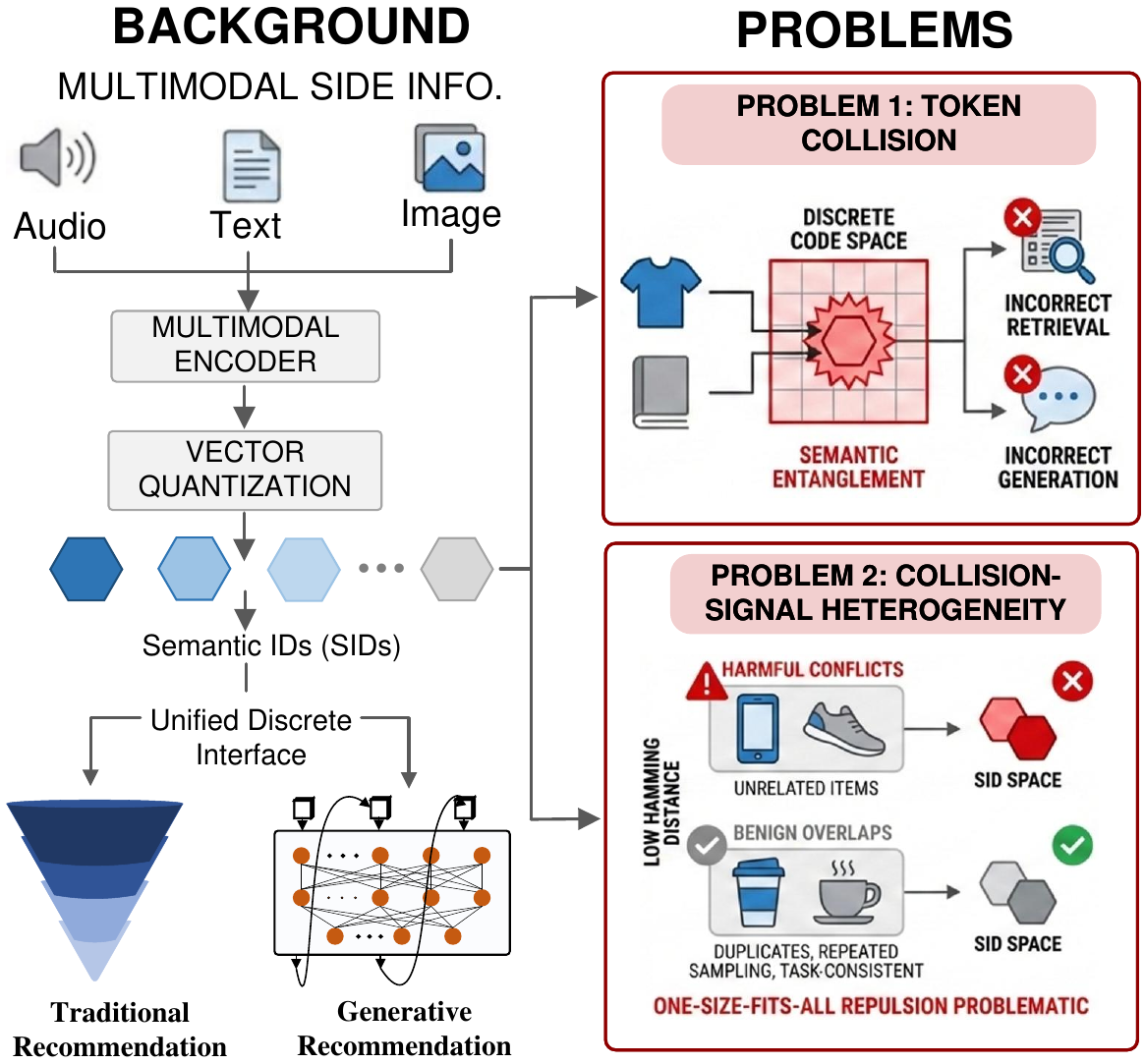}
    \caption{SID learning in large-scale recommendation: multimodal features are encoded and quantized into SIDs as a unified discrete interface for traditional and generative paradigms. Challenges: token collisions cause semantic entanglement, and overlap signals are heterogeneous (harmful vs.\ benign), so one-size-fits-all suppression is problematic.}
    \label{fig:illustration}
\end{figure}

Recent progress in discrete representation learning has established Residual Quantized Variational Autoencoders (RQ-VAE) \cite{DBLP:conf/cvpr/LeeKKCH22} as the mainstream paradigm for mapping multimodal item features into SIDs. By hierarchically quantizing continuous representations into multiple codebooks, RQ-VAE-based methods achieve a favorable balance between expressiveness and compactness, and have demonstrated strong performance across a range of recommendation and generation tasks \cite{DBLP:conf/icde/ZhengHLCZCW24,DBLP:conf/sigir/00050LH0Z25,DBLP:conf/cikm/YeSSWWJ25}. Despite their success, existing SID learning frameworks still exhibit two key limitations that restrict their practical effectiveness \cite{li2025survey}. Figure~\ref{fig:illustration} summarizes the SID pipeline and the two challenges discussed above. First, SID learning suffers from a severe \textbf{collision problem in the discrete space.} When a large corpus of items is compressed into quantized codes, RQ-VAE models often exhibit uneven codebook utilization or centroid collapse during training, causing many semantically unrelated items to be mapped to identical or highly similar SID compositions. Such unintended collisions lead to semantic entanglement, making it difficult for downstream models to distinguish conceptually distinct items. Second, \textbf{collision signals observed during training are highly heterogeneous and require finer-grained qualification}. Apparent SID collisions (i.e., identical or highly overlapping code compositions) may reflect genuinely harmful conflicts between unrelated items, but they can also arise from benign factors such as repeated sampling of the same item or task-consistent relations intentionally introduced by the training pipeline. Without differentiating these cases, a one-size-fits-all collision suppression strategy may inadvertently push apart benign pairs and interfere with downstream alignment.

To address these challenges, we propose \textbf{Qualification-Aware Semantic ID Learning (QuaSID)}, an end-to-end SID learning framework that learns collision-qualified SIDs by applying repulsion only to qualified conflict pairs and calibrating the repulsion strength by collision severity. QuaSID introduces Hamming-guided Margin Repulsion (HaMR), which converts unexpectedly low Hamming distances between SIDs observed during training into explicit geometric margin constraints on the encoder space.
When semantically unrelated items exhibit overly small Hamming distances in their SID compositions, HaMR interprets such cases as insufficient separation in the learned representations and enforces adaptive separation in the continuous embedding space, with penalty strength modulated by collision severity.
To avoid repulsing benign overlaps or structurally biased pairs, QuaSID further employs Conflict-Aware Valid Pair Masking (CVPM), which excludes same-item duplicates and constructed collaborative positives from collision supervision, yielding a cleaner supervision set for repulsion-based learning.
Finally, QuaSID integrates a dual-tower contrastive objective to inject collaborative signals into the tokenization process, which helps align SIDs with downstream recommendation objectives. We conduct extensive offline and online experiments to validate QuaSID.
On public benchmarks, QuaSID improves top-$K$ ranking quality by \textbf{5.9\%} on average over the best baseline while also increasing SID composition diversity.
In a 5-day online A/B test with 5\% traffic on Kuaishou e-commerce, QuaSID-learned SIDs improve ranking GMV-S2 by \textbf{2.38\%} and bring particularly strong gains on cold-start retrieval, with completed orders improving by up to \textbf{6.42\%} on the $100\mathrm{vv}$ cold-start segment.
Moreover, we show that the HaMR repulsion loss is plug-and-play and consistently enhances a wide range of SID learning frameworks across datasets. The main contributions of this work are as follows:
\begin{itemize}
  \item We propose QuaSID, a qualification-aware framework for learning collision-qualified SIDs, and introduce HaMR to translate low-Hamming SID overlaps into severity-aware geometric margin constraints in the encoder space.
  \item We introduce CVPM to qualify collision supervision by masking protocol-induced benign overlaps (e.g., duplicates and constructed positives), yielding a cleaner conflict set for repulsion-based learning.

  \item We validate QuaSID through extensive offline experiments and large-scale online A/B tests, demonstrating consistent improvements in ranking quality, SID composition diversity, and business-critical metrics.
\end{itemize}

\section{Related Work}

\subsection{Vector Quantization for Semantic Tokenization}
Vector quantization (VQ) maps continuous embeddings to a finite set of discrete codewords, typically trained with reconstruction objectives and commitment-style regularizers, and optimized via straight-through or relaxed estimators \cite{DBLP:conf/nips/OordVK17}. To expand capacity without inflating a single vocabulary, modern tokenizers often adopt compositional designs: residual vector quantization (RQ) stacks multiple codebooks to iteratively quantize residuals, achieving strong expressiveness with small per-layer codebooks \cite{DBLP:conf/cvpr/LeeKKCH22,DBLP:journals/taslp/ZeghidourLOST22}. Grouped variants further reduce cross-dimension interference and can improve optimization stability and utilization \cite{DBLP:journals/corr/abs-2305-02765}. Beyond architecture, a line of work improves assignment and normalization, refines gradient estimators, and explicitly diagnoses collapse and under-utilization \cite{zhu2025addressing,DBLP:conf/icml/ChiuQZYW22,DBLP:journals/corr/abs-2106-04283}. More recently, Finite Scalar Quantization (FSQ) replaces explicit codebooks with per-dimension scalar quantization over fixed levels, inducing an implicit Cartesian-product codebook and simplifying training \cite{DBLP:conf/iclr/MentzerMAT24}. Overall, these advances primarily improve quantization capacity and code usage for generic embedding compression.

While the above tokenizers substantially improve quantization capacity and code usage, they are largely developed under a reconstruction oriented paradigm. In Semantic ID construction, discrete codes serve as identifiers and indexing keys for retrieval and ranking, where the key failure mode is confusability between semantically distinct items. Since reconstruction-dominated VQ objectives only weakly constrain such confusability, collision control is typically indirect and can be misaligned with recommendation needs \cite{DBLP:conf/iclr/FiftyJDILATR25,DBLP:conf/iclr/JangGP17}. This gap motivates incorporating recommendation-aligned supervision and discrete-space diagnostics to explicitly shape the encoder and reduce harmful collisions.

\subsection{SID-based Recommendation}
SID-based recommendation represents each item as a fixed-length sequence of discrete semantic tokens, offering compact, reusable, and index-friendly item representations that avoid vocabulary explosion \cite{li2025survey}. This interface has been adopted in both discriminative \cite{DBLP:conf/recsys/ZhengHPRWXNL00L25,DBLP:conf/cikm/LuoCSYHYLZ0HQZZ25,DBLP:conf/recsys/SinghV0KSZHHWTC24,DBLP:conf/kdd/WuTLHMXG25,DBLP:conf/cikm/YeSSWWJ25} and generative recommendation systems \cite{DBLP:journals/corr/abs-2506-13695,DBLP:journals/corr/abs-2509-03236,DBLP:conf/cikm/HanYCJJ0MHLJHZY25}, demonstrating broad applicability across paradigms.

Early SID learning typically follows a projection--quantization--reconstruction pipeline built on VQ/RQ tokenizers \cite{DBLP:conf/nips/OordVK17,DBLP:conf/cvpr/LeeKKCH22}. Recent methods make SID learning more task-aware by injecting collaborative signals into code assignment and training objectives, including VQ-Rec \cite{DBLP:conf/www/HouHMZ23}, LC-Rec \cite{DBLP:conf/icde/ZhengHLCZCW24}, RQ-KMeans \cite{DBLP:journals/corr/abs-2506-13695}, ETEGRec \cite{DBLP:conf/sigir/00050LH0Z25}, and LMIndexer \cite{DBLP:conf/icml/JinZ0CW0W0LLW0T24}. Several works further target collision or assignment imbalance explicitly. SaviorRec \cite{DBLP:journals/corr/abs-2508-01375} integrates Sinkhorn-based entropy-regularized optimal transport to encourage more uniform assignments (see \cite{DBLP:conf/nips/Cuturi13} for the Sinkhorn algorithm); LETTER constrains per-codebook $k$-means by limiting the maximum number of items per centroid \cite{DBLP:conf/cikm/0007BLZ0FNC24}; HiD-VAE adds regularization to penalize excessive code overlap across distinct items \cite{DBLP:journals/corr/abs-2508-04618}. Despite their effectiveness, most collision-aware strategies rely on coarse overlap signals and apply uniform suppression without qualifying what a discrete agreement represents. In large-scale pipelines, low-Hamming overlaps can be heterogeneous, reflecting harmful collisions between unrelated items or benign agreements induced by repeated sampling and task-consistent relations. Indiscriminate suppression can introduce spurious separation and destabilize training, highlighting the need for qualification-aware and severity-adaptive collision control that selects valid conflict pairs and scales repulsion by collision severity.

\section{Preliminaries}

\paragraph{Notations.} Let $\mathcal{U}$ and $\mathcal{I}$ denote the user set and item set, with $|\mathcal{U}|=N_u$ and $|\mathcal{I}|=N_i$. 
Each item $i \in \mathcal{I}$ is associated with a multimodal feature vector 
$\mathbf{x}_i \in \mathbb{R}^{d_{\text{in}}}$, which may include visual, textual, and audio attributes. Let $f_{\theta}(\cdot)$ denote an encoder network that maps the multimodal item feature $\mathbf{x}_i$ into a continuous latent embedding, defined as:
\begin{equation}
\mathbf{z}_i = f_{\theta}(\mathbf{x}_i) \in \mathbb{R}^{d}.
\end{equation}

We adopt a residual vector quantization scheme with $L$ codebooks. 
The $l$-th codebook is denoted as $\mathcal{C}^{(l)} = \{\mathbf{c}^{(l)}_k\}_{k=1}^{K}$, where $K$ is the codebook size. 
For each layer $l$, a discrete index $s_i^{(l)} \in \{1,\dots,K\}$ is selected, producing a quantized embedding 
$\mathbf{q}_i^{(l)} = \mathbf{c}^{(l)}_{s_i^{(l)}}$. The SID of item $i$ is represented as a sequence of discrete tokens:
\begin{equation}
\mathbf{s}_i = \big[s_i^{(1)}, s_i^{(2)}, \dots, s_i^{(L)}\big] \in \{1,\dots,K\}^L.
\end{equation}

\paragraph{Problem definition.} Given a set of multimodal items $\mathcal{I}$ with their feature vectors 
$\{\mathbf{x}_i \mid i \in \mathcal{I}\}$ and a set of observed collaborative item-item pairs 
$\mathcal{P} = \{(i_t, i_p)\}$ derived from user interaction logs, our goal is to learn a SID generator 
\[
g_{\theta}: \mathbf{x}_i \mapsto \mathbf{s}_i \in \{1,\dots,K\}^{L}.
\]
Formally, we aim to jointly learn the encoder $f_{\theta}$, the codebooks ${\mathcal{C}^{(l)}}_{l=1}^{L}$, and the discrete assignment mechanism. The goal is to produce collision-qualified SIDs that remain compatible with collaborative signals and can be deployed in both discriminative and generative recommendation systems.

\begin{figure}
    \centering
    \includegraphics[width=1.0\linewidth]{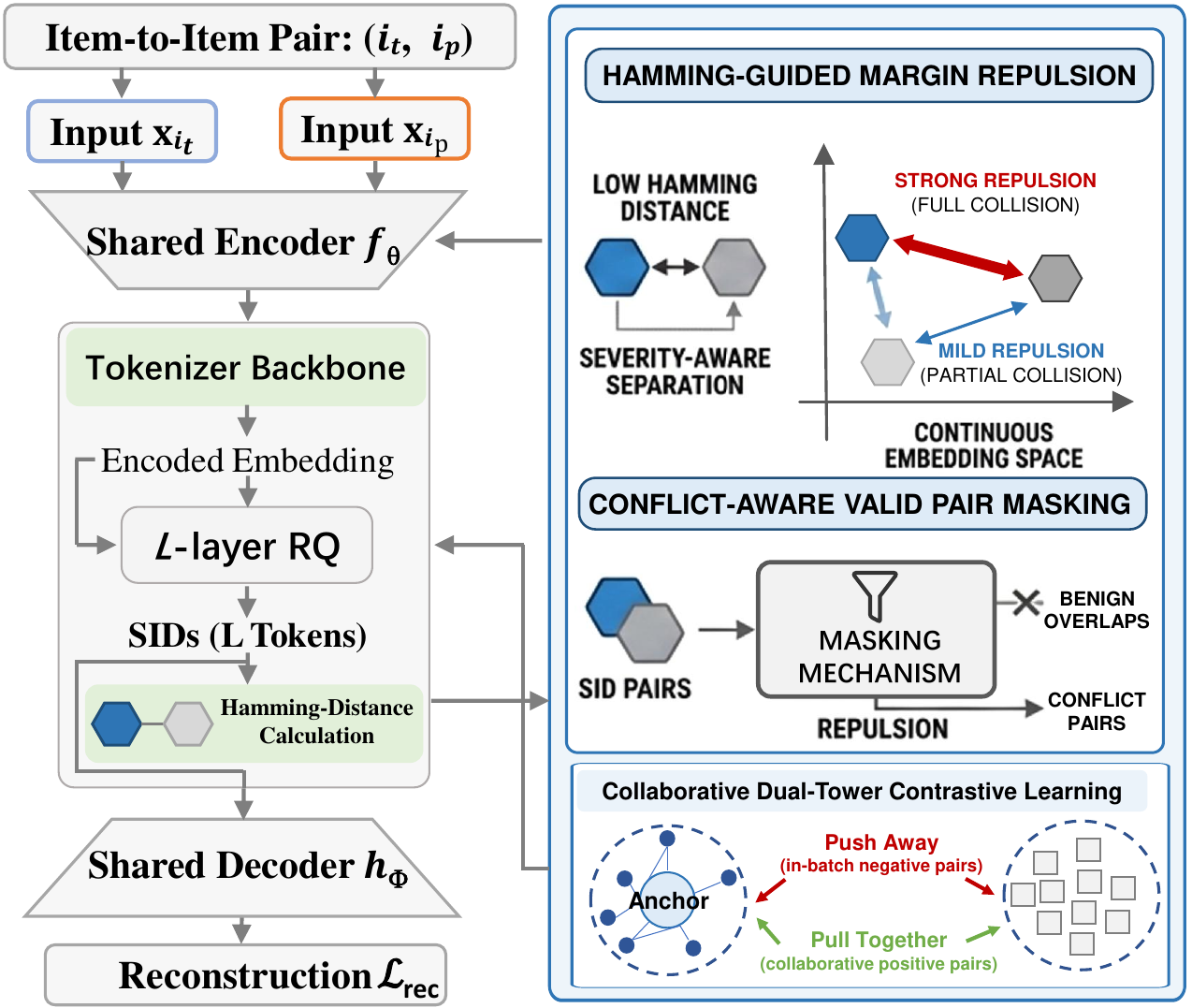}
    \caption{The overall framework of QuaSID.}
    \label{fig:framework}
\end{figure}
\section{Method}


\subsection{Overall Framework}
\paragraph{Tokenizer Backbone.}
As summarized in Figure~\ref{fig:framework}, QuaSID is an end-to-end SID learner that
grounds discrete codes in both multimodal semantics and collaborative signals.
Given a collaborative item-to-item pair,
QuaSID first maps each multimodal input $\mathbf{x}$ into a continuous embedding via a shared encoder $f_{\theta}$.
The embedding is then discretized by an $L$-layer residual vector quantizer to produce an $L$-token SID
$\mathbf{s}_i$ together with its quantized representation $\hat{\mathbf{z}}_i$.
To preserve semantic fidelity under discretization, a decoder $h_{\phi}$ reconstructs the original feature
from $\hat{\mathbf{z}}_i$, yielding a reconstruction-based regularization of the discrete interface.
QuaSID further introduces (i) masked hamming-guided margin repulsion to suppress
harmful in-batch SID collisions, and (ii) collaborative dual-tower contrastive learning to align learned SIDs
with downstream behavioral semantics. Concretely, we encode the trigger and target items with a shared encoder $f_{\theta}$ as follows:
\begin{equation}
\mathbf{z}_{i_t}=f_{\theta}(\mathbf{x}_{i_t}),\qquad
\mathbf{z}_{i_p}=f_{\theta}(\mathbf{x}_{i_p}).
\end{equation}
Here $\mathbf{x}_{i_t},\mathbf{x}_{i_p}\in\mathbb{R}^{d_{\text{in}}}$ are multimodal item features, $\mathbf{z}_{i}\in\mathbb{R}^{d}$ denotes the continuous embedding. For each item $i$, we initialize $\mathbf{r}_i^{(0)}=\mathbf{z}_i$ and iteratively quantize the residual
using $L$ codebooks, yielding the quantized embedding $\hat{\mathbf{z}}_i$ and SID $\mathbf{s}_i$:
\begin{equation}
\label{eq:rq_backbone}
\begin{aligned}
\mathbf{q}_i^{(l)} &= \mathbf{c}^{(l)}_{s_i^{(l)}}, \qquad
\mathbf{r}_i^{(l)} = \mathbf{r}_i^{(l-1)} - \mathbf{q}_i^{(l)}, \qquad l=1,\dots,L,\\
\hat{\mathbf{z}}_i &= \sum_{l=1}^{L}\mathbf{q}_i^{(l)}, \qquad
\mathbf{s}_i = \big[s_i^{(1)},\dots,s_i^{(L)}\big].
\end{aligned}
\end{equation}
We reconstruct the multimodal input from $\hat{\mathbf{z}}_i$ using a decoder $h_{\phi}$: $\hat{\mathbf{x}}_i = h_{\phi}(\hat{\mathbf{z}}_i).$ The backbone is trained with a reconstruction term and the standard residual-quantization objective:
\begin{subequations}
\label{eq:rq_losses}
\begin{align}
\mathcal{L}_{\text{rec}}
&= \frac{1}{|\mathcal{B}|}\sum_{i\in\mathcal{B}}
\ell_{\text{rec}}(\hat{\mathbf{x}}_i,\mathbf{x}_i),
\label{eq:rq_losses_rec}\\
\mathcal{L}_{\text{rq}}
&= \frac{1}{|\mathcal{B}|}\sum_{i\in\mathcal{B}}\sum_{l=1}^{L}
\Big(
\|\mathrm{sg}[\mathbf{r}_i^{(l-1)}]-\mathbf{q}_i^{(l)}\|_2^2
\nonumber\\[-1mm]
&\qquad\qquad
+\beta\|\mathbf{r}_i^{(l-1)}-\mathrm{sg}[\mathbf{q}_i^{(l)}]\|_2^2
\Big),
\label{eq:rq_losses_rq}
\end{align}
\end{subequations}
where $\ell_{\text{rec}}(\cdot,\cdot)$ is instantiated as the $L_2$ loss, $\mathrm{sg}[\cdot]$ denotes
stop-gradient, and $\beta$ controls the commitment strength.

\begin{figure}
    \centering
    \includegraphics[width=0.9\linewidth]{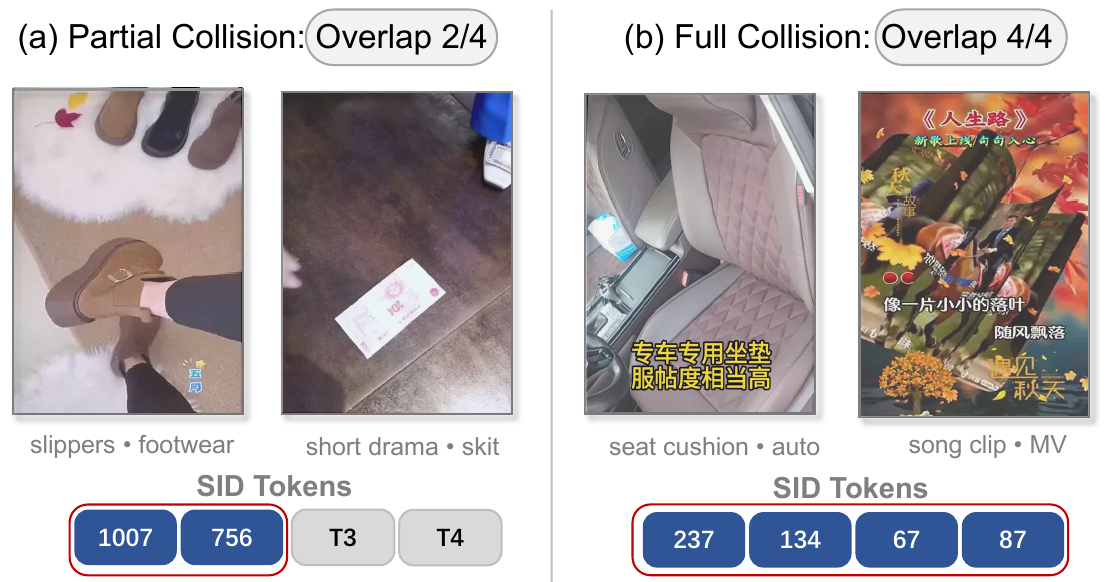}
    \caption{Representative in-batch collision cases in industrial SID training. (a) Partial collision with 2/4 token overlap. (b) Complete collision with identical 4-token SIDs between semantically dissimilar items.}
    \label{fig:collision_case}
\end{figure}

\paragraph{Collaborative Dual-Tower Contrastive Alignment.}
Beyond reconstruction and quantization, QuaSID injects collaborative semantics into tokenization via a dual-tower contrastive objective defined on observed item--item co-engagement pairs (denoted as trigger--target pairs $(i_t,i_p)$). In this work, we adopt Swing~\cite{DBLP:journals/corr/abs-2010-05525} to construct such collaborative pairs, together with data cleaning and filtering strategies to ensure high-quality supervision. 

Given a batch of normalized encoded trigger embeddings $\{\mathbf{e}_t^m\}_{m=1}^B$ and target embeddings $\{\mathbf{e}_p^n\}_{n=1}^B$, we compute the scaled similarities across the mini-batch:
\begin{equation}
\mathbf{S}_{m,n}
= \frac{(\mathbf{e}_t^m)^{\top} \mathbf{e}_p^n}{\tau},
\end{equation}
where $\tau$ is a temperature hyperparameter. The contrastive learning objective follows the InfoNCE formulation:
\begin{equation}
\mathcal{L}_{\text{cl}} 
= - \frac{\lambda_{\text{cl}}}{B} \sum_{m=1}^{B} 
\log \frac{\exp(\mathbf{S}_{m,m})}
{\sum_{n=1}^{B} \mathbb{I}[M_{m,n}=1]\exp(\mathbf{S}_{m,n})},
\end{equation}
where $\lambda_{\text{cl}}$ is the contrastive loss weight and $M_{m,n}$ is a masking indicator.
To reduce false-negative bias, we set $M_{m,n}=0$ for $n\neq m$ when $\mathrm{id}(\mathbf{e}_t^n)=\mathrm{id}(\mathbf{e}_t^m)$, and $M_{m,n}=1$ otherwise.
This task-oriented signal complements semantic reconstruction by encouraging SIDs to preserve behaviorally meaningful proximity.

\subsection{Hamming-guided Margin Repulsion with Conflict-Aware Masking}
\label{sec_hamr}
A core component of QuaSID is \textbf{HaMR}, a mask-aware repulsion mechanism that converts \emph{in-batch low-Hamming SID overlaps} into explicit geometric margin constraints on the encoder space. Under a well-utilized discrete space, it is statistically uncommon for two distinct instances in the same mini-batch to share many SID tokens; hence unusually low Hamming distances within a mini-batch suggest insufficient separation in the learned representations. Figure~\ref{fig:collision_case} presents representative collision cases observed in real-world training scenarios. However, mini-batches also contain overlaps that should \emph{not} be repelled by design, such as duplicate occurrences of the same underlying item and the constructed item--item positives used for the contrastive objective. Therefore, QuaSID employs Conflict-Aware Valid Pair Masking (CVPM) to address the \emph{heterogeneity} of apparent in-batch overlaps by masking these protocol-induced non-conflict pairs before applying repulsion, yielding a cleaner set of candidate conflict pairs for HaMR.

\subsubsection{Conflict-Aware Valid Pair Masking (CVPM)} 

We propose CVPM to \emph{qualify} collision supervision by masking item--item pairs whose discrete overlaps are likely benign, thereby restricting repulsion to a cleaner and more informative subset of candidate conflict pairs. Let the mini-batch be formed by concatenating triggers and targets, yielding $2B$ instances indexed by $i\in\{1,\dots,2B\}$, and let $\mathrm{id}(i)$ denote the underlying item ID of instance $i$.

(1) Collaborative-positive mask: we exclude constructed collaborative positives $(i_t,i_p)$ used by the contrastive objective from collision supervision; otherwise, repulsion would directly contradict task-aligned similarity:
\[
\mathbf{M}^{\text{i2i}}_{ij} =
\begin{cases}
0, & (i \le B \wedge j=i+B) \lor (j \le B \wedge i=j+B), \\
1, & \text{otherwise}.
\end{cases}
\]

(2) Same-item exclusion mask: we exclude pairs that correspond to the same underlying item ID, covering both self-pairs $(i=j)$ and duplicate occurrences caused by repeated sampling. This avoids treating repeated exposures as collisions:
\[
\mathbf{M}^{\text{item}}_{ij} = \mathbb{I}\big[\mathrm{id}(i) \neq \mathrm{id}(j)\big].
\]

Combining the two masks yields a pairwise filter for collision supervision, retaining only the qualified in-batch pairs $(i,j)$:
\begin{equation}
\mathbf{M} = \mathbf{M}^{\text{i2i}} \odot \mathbf{M}^{\text{item}},
\end{equation}
where $\odot$ denotes the element-wise (Hadamard) product. We incorporate $\mathbf{M}$ as a binary gate in the HaMR loss to mask out excluded in-batch pairs $(i,j)$ during repulsion computation.

\subsubsection{Hamming-guided margin repulsion (HaMR)}
We view unexpectedly low-Hamming SID agreements as conflict signals. Accordingly, conditioned on the CVPM valid-pair mask $\mathbf{M}$, we enforce a severity-aware cosine-distance margin between the normalized encoder embeddings to suppress harmful collisions. Let $\mathbf{s}_i\in\{1,\dots,K\}^{L}$ denote the SID and $\mathbf{e}_i = \mathbf{z}_i / \|\mathbf{z}_i\|_2$ the normalized embedding. We define the pairwise Hamming-distance matrix $\mathbf{H}\in\mathbb{N}^{2B\times 2B}$ and
the cosine-distance matrix $\mathbf{D}\in\mathbb{R}^{2B\times 2B}$ by
\begin{align}
\mathbf{H}_{ij} &= d_H(i,j) = \sum_{l=1}^{L} \mathbb{I}[s_i^{(l)} \neq s_j^{(l)}], \\
\mathbf{D}_{ij} &= d_c(i,j) = 1 - \mathbf{e}_i^\top \mathbf{e}_j.
\end{align}

\begin{figure}
    \centering
    \includegraphics[width=1\linewidth]{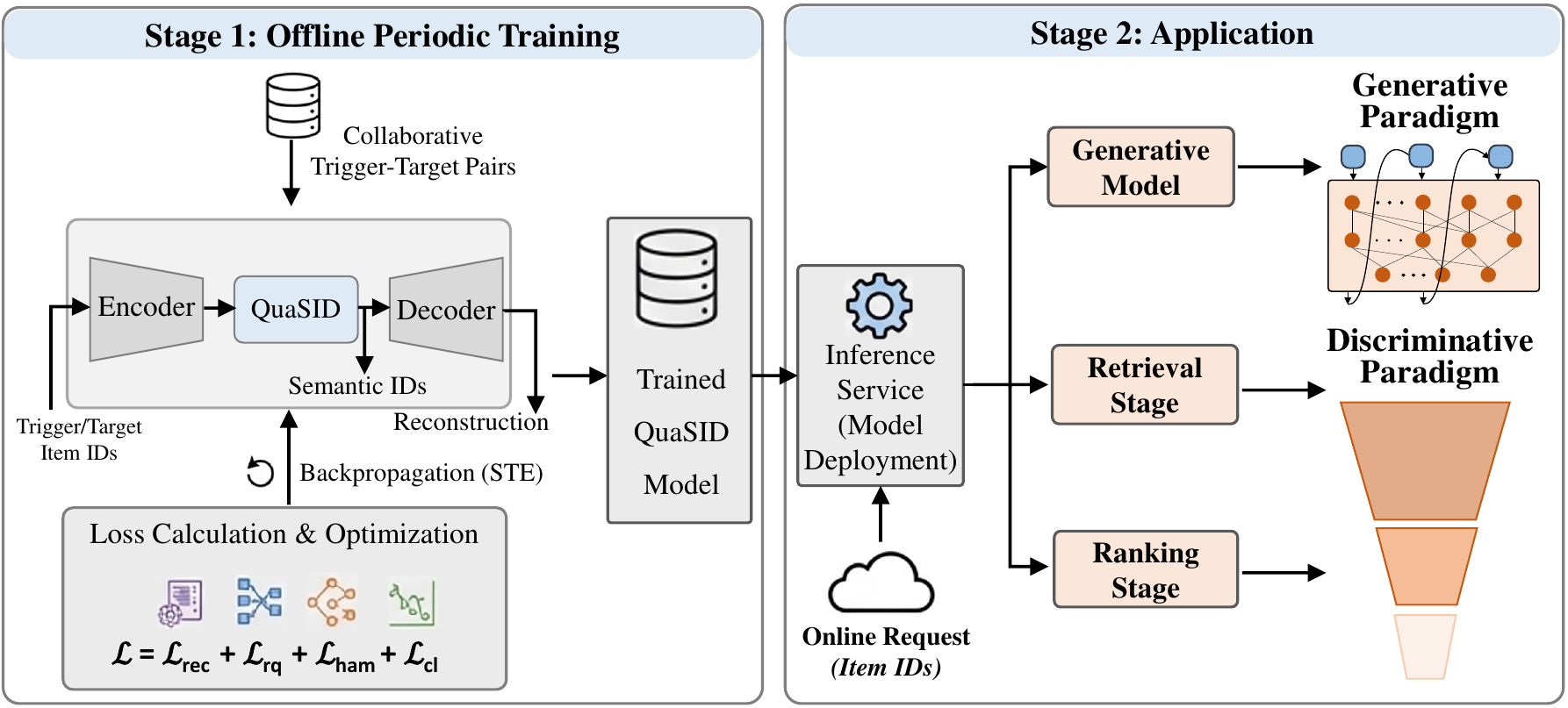}
    \caption{QuaSID deployment and application pipeline.}
    \label{fig:deployment}
\end{figure}

Based on the Hamming distance, we distinguish between two types of collisions: \emph{full collision} and \emph{partial collision}. This distinction allows us to adaptively control the penalty strength according to the severity of the collision. Intuitively, a full collision represents a more problematic case and thus warrants a stronger penalty. Using the masks defined above, we partition candidate pairs into two sets:
\begin{align*}
\Omega_{\text{full}} &= \big\{(i,j)\mid \mathbf{H}_{ij}=0 \text{ and } \mathbf{M}_{ij}=1\big\}, \\
\Omega_{\text{partial}} &= \big\{(i,j)\mid 0<\mathbf{H}_{ij}\le R \text{ and } \mathbf{M}_{ij}=1\big\},
\end{align*}
where $R$ is a hyperparameter that determines the Hamming radius for considering partial collisions. For each pair $(i,j)$ we penalize small cosine distances via margin-based hinge losses:
\begin{align}
\mathcal{L}_{\text{full}}(i,j) &= \max\big(0, m_{\text{full}} - \mathbf{D}_{ij}\big),\\
\mathcal{L}_{\text{partial}}(i,j) &= \max\big(0, m_{\text{partial}} - \mathbf{D}_{ij}\big),
\end{align}
where $m_{\text{full}}$ and $m_{\text{partial}}$ are predefined margins with $m_{\text{full}}\ge m_{\text{partial}}$. The masked hamming-guided margin is therefore computed as:
\begin{equation}
\begin{split}
\mathcal{L}_{\text{HaMR}}
&= \frac{\lambda_{\text{full}}}{|\Omega_{\text{full}}|+\epsilon}
   \sum_{(i,j)\in\Omega_{\text{full}}}\mathcal{L}_{\text{full}}(i,j) \\
&\quad + \frac{\lambda_{\text{partial}}}{|\Omega_{\text{partial}}|+\epsilon}
   \sum_{(i,j)\in\Omega_{\text{partial}}}\mathcal{L}_{\text{partial}}(i,j),
\end{split}
\label{eq:ham_masked}
\end{equation}
where $\lambda_{\text{full}}$ and $\lambda_{\text{partial}}$ are weighting coefficients and
$\epsilon>0$ is a small constant for numerical stability.

Eq.~\eqref{eq:ham_masked} imposes geometric constraints on the encoder space for the selected pairs: for any $(i,j)\in\Omega_{\text{full}}$ (resp. $\Omega_{\text{partial}}$), achieving zero hinge loss implies $d_c(i,j)\ge m_{\text{full}}$ (resp. $d_c(i,j)\ge m_{\text{partial}}$). With CVPM filtering constructed positives and same-ID pairs, repulsion focuses on discretely similar item pairs that are unlikely to be benign artifacts, increasing their angular separation before quantization. Over training, such targeted separation reshapes the continuous embedding space and reduces collision frequency, without enforcing uniform repulsion across all item pairs.

\subsection{System Deployment and Application}
QuaSID has been successfully deployed and applied on Kuaishou’s recommendation system, as illustrated in Figure~\ref{fig:deployment}. The deployment and application pipeline consists of the following key stages: (1) Offline periodic training: Based on the production speed of the trigger-target pair, we train our model periodically using the overall optimization objective:
\begin{equation}
\mathcal{L} 
= \mathcal{L}_{\text{rec}} 
+ \mathcal{L}_{\text{rq}} 
+ \mathcal{L}_{\text{HaMR}} 
+ \mathcal{L}_{\text{cl}}.
\end{equation}
All components are trained end-to-end using backpropagation, where the discrete code assignments are optimized via the straight-through estimator (STE) to enable differentiable learning. (2) Application for retrieval and ranking stage: The trained QuaSID is deployed as an inference service to project item ids into SIDs. QuaSID-learned SIDs correspond to a new lookup table and are used in both retrieval and ranking stages. The retrieval stage including conventional retrieval and SID-based generative retrieval. At the ranking stage, SIDs are used to derive cross features and matching side features, serving as lightweight semantic signals that augment the existing feature set.

\begin{table}[t]
\centering
\caption{Dataset statistics used in the experiments.}
\label{tab:dataset_statistics}
\renewcommand{\arraystretch}{0.9} 
\begin{tabular}{lccc}
\toprule
Dataset & \#Users & \#Items & \#Interactions \\
\midrule
Amazon-Beauty & 22{,}363 & 12{,}101 & 1{,}048{,}296 \\
Amazon-Toys   & 19{,}412 & 11{,}924 &   905{,}253 \\
\bottomrule
\end{tabular}
\end{table}

\begin{table*}[t]
\centering
\caption{Performance comparison on Amazon-Beauty and Amazon-Toys. Bold denotes the best result, and underlined denotes the second-best result.}
\label{tab:tokenizer_comparison}
\resizebox{\textwidth}{!}{
\begin{tabular}{l|ccccc|ccccc}
\hline
 & \multicolumn{5}{c|}{\textbf{Amazon-Beauty}} & \multicolumn{5}{c}{\textbf{Amazon-Toys}} \\
\textbf{Tokenizer} 
& HR@5 & HR@10 & NDCG@5 & NDCG@10 & Entropy
& HR@5 & HR@10 & NDCG@5 & NDCG@10 & Entropy \\
\hline
RQ-VAE 
& 0.0225 & 0.0300 & 0.0171 & 0.0195 & 9.3075
& 0.0206 & 0.0256 & 0.0164 & 0.0180 & 9.3068 \\

Improved VQGAN 
& 0.0232 & 0.0304 & 0.0167 & 0.0189 & 9.3569
& 0.0196 & 0.0245 & 0.0157 & 0.0172 & 9.3313 \\

GRVQ 
& 0.0222 & 0.0302 & 0.0161 & 0.0187 & 9.2755
& 0.0195 & 0.0242 & 0.0158 & 0.0173 & 9.2101 \\

RQ-OPQ 
& 0.0205 & 0.0271 & 0.0159 & 0.0180 & 9.3368
& 0.0228 & 0.0278 & 0.0179 & 0.0195 & 9.3521 \\

RQ-VAE-Rotation 
& 0.0236 & 0.0308 & \underline{0.0175} & 0.0198 & 9.3455
& 0.0203 & 0.0257 & 0.0167 & 0.0187 & 9.3290 \\

SimRQ 
& 0.0231 & 0.0297 & 0.0172 & 0.0193 & 9.3526
& 0.0220 & 0.0279 & 0.0171 & 0.0190 & \underline{9.3688} \\

RQ-Kmeans 
& \underline{0.0254} & \underline{0.0379} & 0.0171 & \underline{0.0211} & \underline{9.3793}
& \underline{0.0260} & \underline{0.0347} & \underline{0.0190} & \underline{0.0213} & 9.3460 \\

\hline
QuaSID 
& \textbf{0.0277} & \textbf{0.0392} & \textbf{0.0193} & \textbf{0.0230} & \textbf{9.3901}
& \textbf{0.0266} & \textbf{0.0366} & \textbf{0.0193} & \textbf{0.0225} & \textbf{9.3794} \\
\hline
\end{tabular}
}
\end{table*}

\begin{table*}[t]
\centering
\small
\setlength{\tabcolsep}{3pt}
\caption{Effect of HaMR ($\mathcal{L}_{\text{HaMR}}$) on different semantic ID learning methods.}
\label{tab:ham_generalization}
\begin{tabular}{llllllllllll}
\toprule
 & \multicolumn{6}{c}{\textbf{Amazon-Beauty}} & \multicolumn{5}{c}{\textbf{Amazon-Toys}} \\
\cmidrule(lr){2-7} \cmidrule(lr){8-12}
\textbf{Tokenizer} 
& HR@5 & HR@10 & NDCG@5 & NDCG@10 & Entropy 
&  & HR@5 & HR@10 & NDCG@5 & NDCG@10 & Entropy \\
\midrule
RQ-VAE
& 0.0225 & 0.0300 & 0.0171 & 0.0195 & 9.3075
&  & 0.0206 & 0.0256 & 0.0164 & 0.0180 & 9.3068 \\
\quad + $\mathcal{L}_{\text{HaMR}}$
& 0.0246$_{\textcolor{red}{\uparrow 9.3\%}}$ & 0.0348$_{\textcolor{red}{\uparrow 16.0\%}}$ & 0.0182$_{\textcolor{red}{\uparrow 6.4\%}}$ & 0.0215$_{\textcolor{red}{\uparrow 10.3\%}}$ & 9.3971$_{\textcolor{red}{\uparrow 1.0\%}}$
&  & 0.0239$_{\textcolor{red}{\uparrow 16.0\%}}$ & 0.0290$_{\textcolor{red}{\uparrow 13.3\%}}$ & 0.0187$_{\textcolor{red}{\uparrow 14.0\%}}$ & 0.0203$_{\textcolor{red}{\uparrow 12.8\%}}$ & 9.3795$_{\textcolor{red}{\uparrow 0.8\%}}$ \\
\midrule
Improved VQGAN
& 0.0232 & 0.0304 & 0.0167 & 0.0189 & 9.3569
&  & 0.0196 & 0.0245 & 0.0157 & 0.0172 & 9.3313 \\
\quad + $\mathcal{L}_{\text{HaMR}}$
& 0.0257$_{\textcolor{red}{\uparrow 10.8\%}}$ & 0.0340$_{\textcolor{red}{\uparrow 11.8\%}}$ & 0.0192$_{\textcolor{red}{\uparrow 15.0\%}}$ & 0.0218$_{\textcolor{red}{\uparrow 15.3\%}}$ & 9.3974$_{\textcolor{red}{\uparrow 0.4\%}}$
&  & 0.0239$_{\textcolor{red}{\uparrow 21.8\%}}$ & 0.0301$_{\textcolor{red}{\uparrow 22.8\%}}$ & 0.0187$_{\textcolor{red}{\uparrow 18.9\%}}$ & 0.0207$_{\textcolor{red}{\uparrow 20.1\%}}$ & 9.3797$_{\textcolor{red}{\uparrow 0.5\%}}$ \\
\midrule
GRVQ
& 0.0222 & 0.0302 & 0.0161 & 0.0187 & 9.2755
&  & 0.0195 & 0.0242 & 0.0158 & 0.0173 & 9.2101 \\
\quad + $\mathcal{L}_{\text{HaMR}}$
& 0.0251$_{\textcolor{red}{\uparrow 13.1\%}}$ & 0.0341$_{\textcolor{red}{\uparrow 12.9\%}}$ & 0.0181$_{\textcolor{red}{\uparrow 12.4\%}}$ & 0.0210$_{\textcolor{red}{\uparrow 12.3\%}}$ & 9.3986$_{\textcolor{red}{\uparrow 1.3\%}}$
&  & 0.0232$_{\textcolor{red}{\uparrow 19.0\%}}$ & 0.0284$_{\textcolor{red}{\uparrow 17.4\%}}$ & 0.0186$_{\textcolor{red}{\uparrow 17.7\%}}$ & 0.0202$_{\textcolor{red}{\uparrow 16.8\%}}$ & 9.3808$_{\textcolor{red}{\uparrow 1.9\%}}$ \\
\midrule
RQ-OPQ
& 0.0205 & 0.0271 & 0.0159 & 0.0180 & 9.3368
&  & 0.0228 & 0.0278 & 0.0179 & 0.0195 & 9.3521 \\
\quad + $\mathcal{L}_{\text{HaMR}}$
& 0.0241$_{\textcolor{red}{\uparrow 17.6\%}}$ & 0.0317$_{\textcolor{red}{\uparrow 17.0\%}}$ & 0.0179$_{\textcolor{red}{\uparrow 12.5\%}}$ & 0.0203$_{\textcolor{red}{\uparrow 12.5\%}}$ & 9.3973$_{\textcolor{red}{\uparrow 0.6\%}}$
&  & 0.0227$_{\textcolor{green!60!black}{\downarrow 0.4\%}}$ & 0.0280$_{\textcolor{red}{\uparrow 0.7\%}}$ & 0.0180$_{\textcolor{red}{\uparrow 0.6\%}}$ & 0.0197$_{\textcolor{red}{\uparrow 1.0\%}}$ & 9.3798$_{\textcolor{red}{\uparrow 0.3\%}}$ \\
\midrule
RQ-VAE-Rotation
& 0.0236 & 0.0308 & 0.0175 & 0.0198 & 9.3455
&  & 0.0203 & 0.0257 & 0.0167 & 0.0187 & 9.3290 \\
\quad + $\mathcal{L}_{\text{HaMR}}$
& 0.0242$_{\textcolor{red}{\uparrow 2.5\%}}$ & 0.0327$_{\textcolor{red}{\uparrow 6.2\%}}$ & 0.0175 & 0.0202$_{\textcolor{red}{\uparrow 2.0\%}}$ & 9.3939$_{\textcolor{red}{\uparrow 0.5\%}}$
&  & 0.0219$_{\textcolor{red}{\uparrow 7.9\%}}$ & 0.0272$_{\textcolor{red}{\uparrow 5.8\%}}$ & 0.0174$_{\textcolor{red}{\uparrow 3.9\%}}$ & 0.0191$_{\textcolor{red}{\uparrow 2.0\%}}$ & 9.3765$_{\textcolor{red}{\uparrow 0.5\%}}$ \\
\midrule
SimRQ
& 0.0231 & 0.0297 & 0.0172 & 0.0193 & 9.3526
&  & 0.0220 & 0.0279 & 0.0171 & 0.0190 & 9.3688 \\
\quad + $\mathcal{L}_{\text{HaMR}}$
& 0.0238$_{\textcolor{red}{\uparrow 3.0\%}}$ & 0.0323$_{\textcolor{red}{\uparrow 8.8\%}}$ & 0.0174$_{\textcolor{red}{\uparrow 1.2\%}}$ & 0.0201$_{\textcolor{red}{\uparrow 4.1\%}}$ & 9.3977$_{\textcolor{red}{\uparrow 0.5\%}}$
&  & 0.0227$_{\textcolor{red}{\uparrow 3.2\%}}$ & 0.0283$_{\textcolor{red}{\uparrow 1.4\%}}$ & 0.0182$_{\textcolor{red}{\uparrow 6.4\%}}$ & 0.0200$_{\textcolor{red}{\uparrow 5.3\%}}$ & 9.3793$_{\textcolor{red}{\uparrow 0.1\%}}$ \\
\bottomrule
\end{tabular}
\end{table*}

\section{Experiment}

\subsection{Experimental Setup} 
\subsubsection{Datasets.} We use two public benchmarks for offline evaluation and an industrial dataset for online A/B tests, as detailed below. \textbf{Public Datasets.} For offline generative recommendation evaluation, we use the widely adopted Amazon 2018 review datasets\footnote{https://nijianmo.github.io/amazon/}. Specifically, we select the \emph{Beauty} and \emph{Toys \& Games} subsets to evaluate our method. Following prior work~\cite{DBLP:conf/recsys/Geng0FGZ22}, we apply a standard \emph{5-core} filtering procedure, discarding users and items with fewer than five interactions. We then adopt a leave-one-out strategy to split each dataset into training, validation, and test sets. Item textual fields include \emph{Title}, \emph{Brand}, \emph{Categories}, and \emph{Price}. We concatenate these fields and extract semantic item embeddings using Sentence-T5-XXL~\cite{DBLP:conf/acl/NiACMHCY22}. The dataset statistics are shown in Table~\ref{tab:dataset_statistics}. \textbf{Industrial Dataset.} On an industrial dataset from the Kuaishou platform, we perform online A/B tests to evaluate QuaSID in both generative retrieval and discriminative pipelines (including retrieval and ranking). Each item is associated with rich multimodal side information, including textual descriptions, automatic speech recognition (ASR) transcripts, and key-frame images. Multimodal item embeddings are extracted using multimodal large language models.

\subsubsection{Baselines.}
For offline experiments, we compare QuaSID with strong VQ-based tokenizers, including \textbf{RQ-VAE}~\cite{DBLP:conf/cvpr/LeeKKCH22}, \textbf{GRVQ}~\cite{DBLP:journals/corr/abs-2305-02765}, \textbf{Improved VQGAN}~\cite{DBLP:conf/iclr/YuLKZPQKXBW22}, \textbf{RQ-VAE-Rotation}~\cite{DBLP:conf/iclr/FiftyJDILATR25}, \textbf{SimRQ}~\cite{zhu2025addressing}, \textbf{RQ-OPQ}~\cite{DBLP:journals/corr/abs-2509-03236}, and \textbf{RQ-Kmeans}~\cite{DBLP:conf/cikm/LuoCSYHYLZ0HQZZ25}. Due to space, detailed baseline descriptions are deferred to Appendix~\ref{app:baselines}.

\subsubsection{Evaluation Metrics.}
For offline evaluation on public datasets, we report HitRate@$K$ and NDCG@$K$ with $K \in \{5,10\}$ as the ranking metrics. We additionally report the entropy of \emph{SID compositions} to measure the diversity of discrete ID assignments. 
Let $p(\mathbf{s})$ denote the empirical frequency of an SID sequence $\mathbf{s}$ over the item corpus; we compute
$\mathcal{E}_{\text{SID}} = -\sum_{\mathbf{s}} p(\mathbf{s}) \log p(\mathbf{s})$, where larger values indicate more diverse SIDs (i.e., fewer duplicated compositions). For online A/B tests on the industrial dataset, we report business metrics as described in Section~\ref{ab_test}.

\subsubsection{Implementation Details.}
For offline experiments, we employ TIGER~\cite{DBLP:conf/nips/RajputMSKVHHT0S23} as the generative recommendation backbone for all methods, and keep identical model configurations across QuaSID and baselines. For the codebook setting, we set $L=3$ and $K=256$ in offline experiments, while in industrial experiments we use $L=4$ and $K=1024$. All results are averaged over five random seeds. Additional training details and hyperparameter ranges are provided in Appendix~\ref{app:impl}.

\subsection{Overall Performance}
We compare QuaSID with multiple baselines on the public datasets under the offline generative recommendation setting. The results are shown in Table~\ref{tab:tokenizer_comparison}. All improvements are statistically significant ($p$<0.05). Based on the results, we have the following observations:
\begin{itemize}
    \item As shown in Table~\ref{tab:tokenizer_comparison}, QuaSID consistently achieves the best performance on all ranking metrics (HR$@K$ and NDCG$@K$) across both Amazon-Beauty and Amazon-Toys. The improvements are substantial and consistent compared to strong VQ-based baselines, indicating that collision-reduced and collaboration-aware SID learning translates into superior downstream recommendation quality.

    \item QuaSID also attains the highest entropy on both datasets, indicating more diverse discrete ID assignments and better utilization of the discrete SID space. Together with the ranking improvements, this indicates that the proposed framework enhances task performance while maintaining a compact and efficient discrete interface; by integrating task-aligned supervision with collision-aware regularization, QuaSID jointly improves SID discriminability, task alignment, and SID assignment diversity within a single end-to-end learning objective.

\end{itemize}
\paragraph{Entropy--ranking quality relationship.}
$\mathcal{E}_{\text{SID}}$ measures the diversity of SID compositions (i.e., how evenly the discrete space is utilized).
Across tokenizers, higher $\mathcal{E}_{\text{SID}}$ generally coincides with better ranking metrics, and the overall correlation is positive when aggregating Amazon-Beauty and Amazon-Toys (Pearson $r$ around 0.65, Spearman $\rho$ around 0.72; statistically significant). However, entropy is not a sufficient predictor of ranking quality.
A higher $H_{\text{SID}}$ mainly reflects fewer duplicated compositions, while HR/NDCG further depends on whether the discrete interface preserves task-relevant semantics.
For example, on Amazon-Toys, SimRQ attains high entropy (9.3688, second-best) but underperforms RQ-Kmeans on NDCG@10 (0.0190 vs.\ 0.0213) despite RQ-Kmeans having lower entropy (9.3460).

\begin{figure*}[htbp]
    \centering
    \begin{subfigure}[b]{0.3\textwidth}
        \centering
        \includegraphics[width=\linewidth]{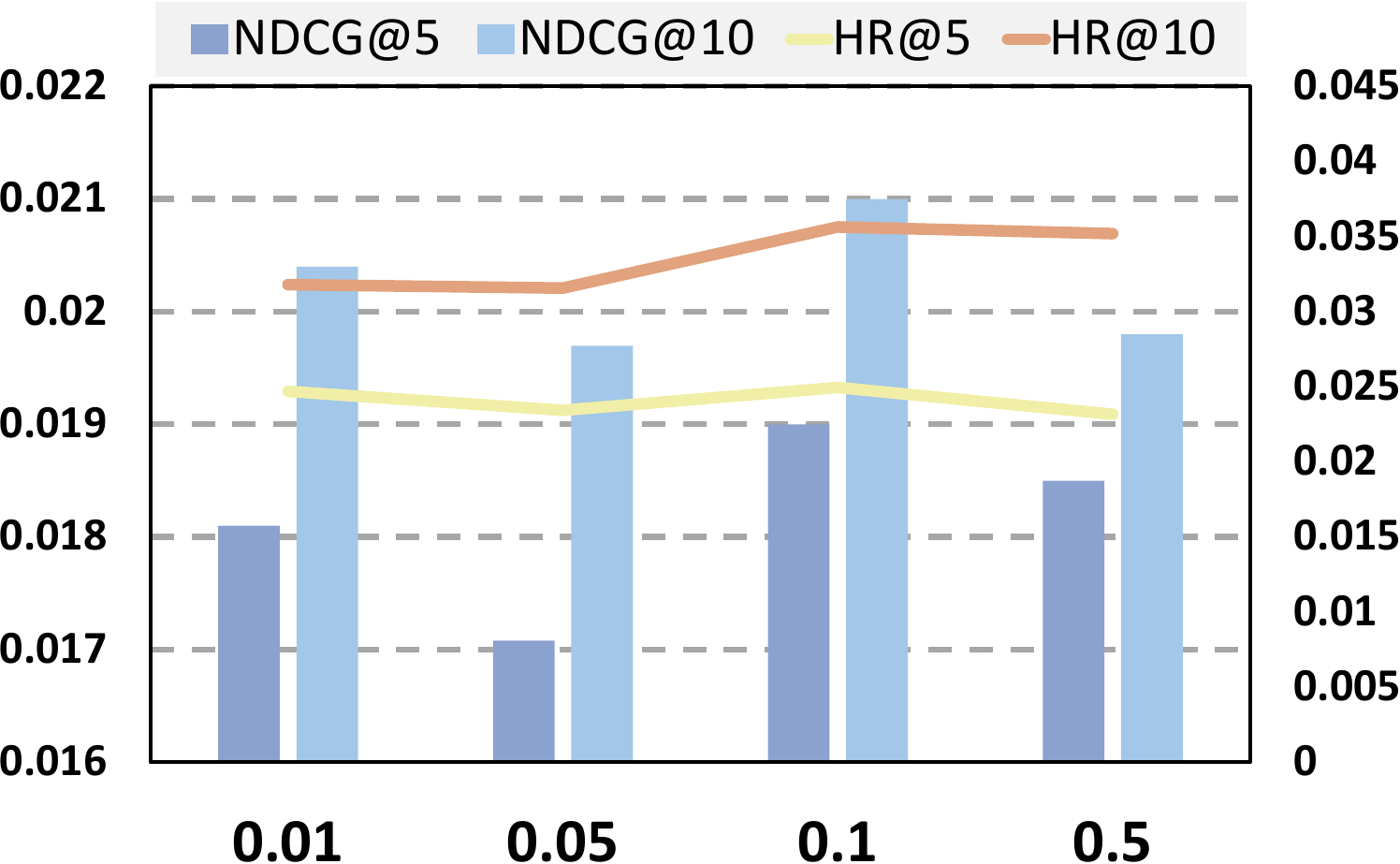}
        \caption{$\lambda_{cl}$}
        \label{fig:sub1}
    \end{subfigure}
    \hfill
    \begin{subfigure}[b]{0.3\textwidth}
        \centering
        \includegraphics[width=\linewidth]{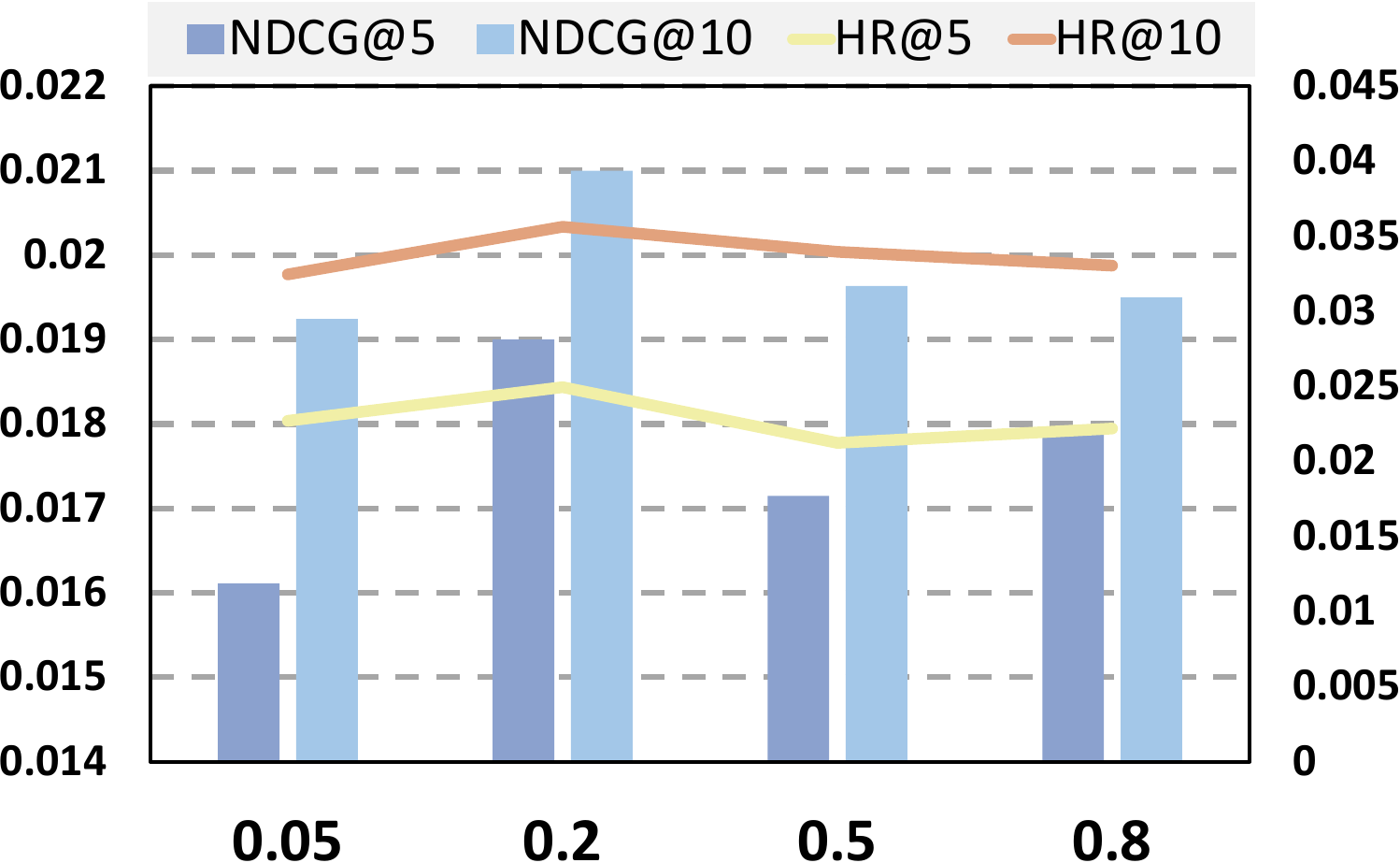}
        \caption{$\lambda_{full}$}
        \label{fig:sub2}
    \end{subfigure}
    \hfill
    \begin{subfigure}[b]{0.3\textwidth}
        \centering
        \includegraphics[width=\linewidth]{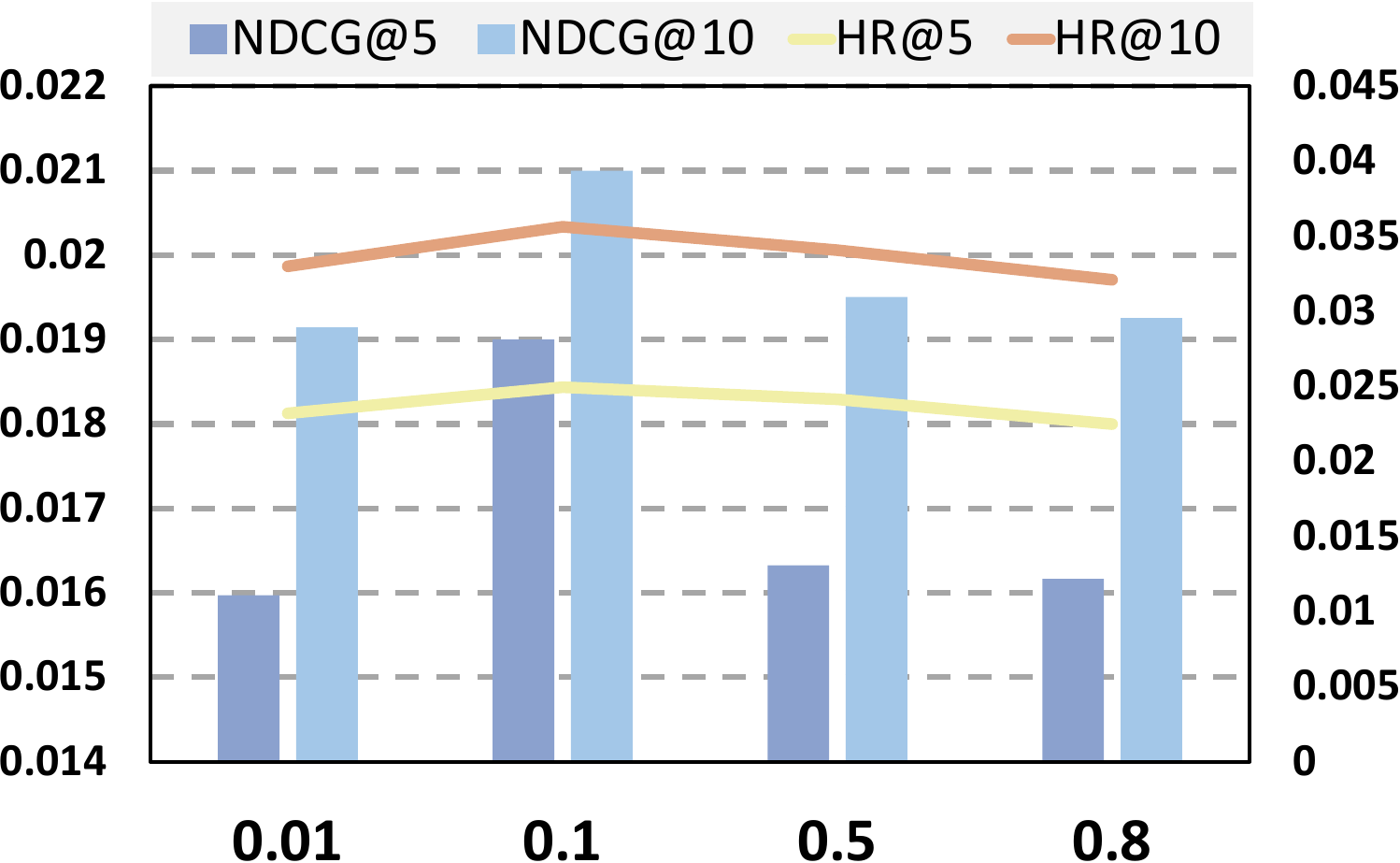}
        \caption{$\lambda_{partial}$}
        \label{fig:sub3}
    \end{subfigure}
    \caption{Hyperparameter sensitivity analysis of QuaSID on the Amazon-Beauty dataset. The bars indicate NDCG@K (left y-axis), while the curves denote HR@K (right y-axis). In each subfigure, one hyperparameter is varied while all others are fixed to their default values.}
    \label{fig:hp_sensitive}
\end{figure*}

\begin{table}[t]
\centering
\caption{Online A/B test results on Kuaishou e-commerce. Relative uplifts (\%) brought by integrating QuaSID-learned SIDs into retrieval and ranking.}
\label{tab:ab_tst}
\small            
\setlength{\tabcolsep}{6pt} 
\begin{tabular}{lccc}
\toprule
\textbf{Setting} 
& \textbf{Completed Orders} 
& \textbf{GMV-S1} 
& \textbf{GMV-S2}\\
\cmidrule(r){2-2}\cmidrule(r){3-3}\cmidrule(r){4-4}
Generative Retrieval
& +0.21\% 
& +1.03\% 
& +0.55\%\\
\cdashline{1-4}
Ranking
& +0.20\% 
& +1.44\% 
& +2.38\% \\
\midrule
& \textbf{Completed Orders} 
& \textbf{GMV} 
& \textbf{GPM} \\
\cmidrule(r){2-2}\cmidrule(r){3-3}\cmidrule(r){4-4}
Retrieval
& +1.09\% 
& +1.69\% 
& +3.25\% \\
Retrieval$_{100vv}$
& +6.42\% 
& +4.67\% 
& +0.21\% \\
Retrieval$_{600vv}$
& +4.69\% 
& +3.11\% 
& +0.53\% \\
\cdashline{1-4}
Ranking$_{100vv}$
& +1.77\% 
& +4.10\% 
& +2.99\% \\
Ranking$_{600vv}$
& +2.64\% 
& +3.88\% 
& +2.78\% \\
\bottomrule
\end{tabular}
\end{table}

\subsection{Plug-and-Play Analysis of HaMR}
To evaluate the generalization ability of the proposed HaMR loss ($\mathcal{L}_{\text{HaMR}}$), we conduct a controlled study where all training settings and hyperparameters are kept identical, and we add the HaMR loss as an auxiliary objective to a diverse set of semantic ID learning baselines. The results are summarized in Table~\ref{tab:ham_generalization}. Note that RQ-Kmeans is excluded from Table~\ref{tab:ham_generalization}, as its tokenizer is not trained under the same end-to-end, gradient-based optimization framework as the tokenizers evaluated in this study. Based on the empirical evidence, we draw the following observations:
\begin{itemize}
    \item As shown in Table~\ref{tab:ham_generalization}, adding the HaMR loss improves the downstream ranking metric NDCG@10 by up to 15.3\% (Beauty) and 20.1\% (Toys), while also increasing entropy by around 0.1\%–1.9\%. This demonstrates that the proposed hamming-guided margin repulsion is model-agnostic and can be seamlessly integrated as a plug-and-play component to enhance semantic ID discriminability, collision reduction, and SID assignment diversity simultaneously.

    \item By comparing Table~\ref{tab:ham_generalization} with the overall performance of QuaSID in Table~\ref{tab:tokenizer_comparison}, we observe a contrast: while adding $\mathcal{L}_{\text{HaMR}}$ to baselines often results in higher entropy than QuaSID, their ranking metrics remain consistently inferior to QuaSID. This suggests that although hamming-guided repulsion is highly effective in improving SID-composition diversity and SID structure, it alone is insufficient to fully optimize downstream recommendation performance in the offline generative setting. The additional gains achieved by QuaSID can be likely related to its task-aware training module, which aligns semantic IDs with recommendation objectives. Together, these results highlight that $\mathcal{L}_{\text{HaMR}}$ and contrastive task supervision are complementary: the former improves discrete representation quality, while the latter ensures effective downstream task alignment.

\end{itemize}

\subsection{Online A/B Test}
\label{ab_test}
We further evaluate QuaSID through large-scale online A/B testing in the e-commerce recommendation scenario of the Kuaishou platform. Approximately $5\%$ of online traffic (covering over 20 million users) is randomly assigned to the treatment group for $5$ days. We report Completed Orders (CO), GMV (including scenario-specific GMV, i.e., GMV-S1/GMV-S2), and GPM (GMV per mille exposures) in Table~\ref{tab:ab_tst}; $100vv$ and $600vv$ denote cold-start videos with fewer than 100 and 600 views, respectively, within the first 48 hours. All results are statistically significant. 

\paragraph{Retrieval.}
We evaluate both \textbf{generative retrieval} and the \textbf{conventional discriminative retrieval} pipeline, together with their cold-start subsets. As shown in Table~\ref{tab:ab_tst}, replacing the tokenizer with QuaSID-learned SIDs improves CO and GMV-related metrics across settings, with substantially larger gains under sparse feedback. In generative retrieval, QuaSID yields +0.21\% CO and +1.03\% / +0.55\% GMV-S1/GMV-S2. The effect is amplified on cold-start traffic, where CO increases by +6.42\% ($100\mathrm{vv}$) and +4.69\% ($600\mathrm{vv}$). GPM follows the same trend.

\paragraph{Ranking.}
QuaSID also delivers stable improvements when integrated into \textbf{discriminative ranking models}. We observe +2.38\% GMV-S2, accompanied by +1.44\% GMV-S1 and +0.20\% CO. While the absolute uplifts are smaller than those in retrieval, the gains remain consistently positive across scenarios, and cold-start segments still benefit noticeably. Overall, these results demonstrate that QuaSID-learned SIDs can be deployed in production ranking and retrieval stacks and translate into measurable business impact.

\begin{table}[t]
\centering
\caption{Ablation study results on Amazon-Beauty and Amazon-Toys datasets.}
\label{tab:ablation}
\renewcommand{\arraystretch}{0.9} 
\resizebox{\linewidth}{!}{
\begin{tabular}{lcccc}
\toprule
\multirow{2}{*}{Tokenizer} 
& \multicolumn{2}{c}{Amazon-Beauty} 
& \multicolumn{2}{c}{Amazon-Toys} \\
\cmidrule(lr){2-3} \cmidrule(lr){4-5}
& HR@5 & NDCG@5 & HR@5 & NDCG@5 \\
\midrule
QuaSID        & 0.0277 & 0.0193 & 0.0266 & 0.0193 \\
w/o $CVPM$      & 0.0263 & 0.0181 & 0.0264 & 0.0190 \\
w/o $HaMR$     & 0.0254 & 0.0170 & 0.0261 & 0.0189 \\
\bottomrule
\end{tabular}
}
\end{table}

\subsection{Ablation Study}
We conduct ablation studies on the Amazon-Beauty and Amazon-Toys datasets to examine the effectiveness of the key components in QuaSID. Specifically, we evaluate: (i) \emph{CVPM}, which qualifies collision supervision by masking pairs that are likely to reflect benign overlaps; (ii) the hamming-guided margin repulsion loss $\mathcal{L}_{HaMR}$ for collision suppression. Results are reported in Table~\ref{tab:ablation}.

\paragraph{Effect of CVPM.}
To assess whether valid-pair masking is necessary for reliable collision supervision, we remove CVPM while keeping the remaining objectives unchanged (denoted as \emph{w/o CVPM}). As shown in Table~\ref{tab:ablation}, this variant consistently degrades performance on both datasets, with a more noticeable drop on Amazon-Beauty. This supports our motivation that apparent in-batch collisions are heterogeneous and require qualification. Without CVPM, benign overlaps can be mistakenly included in the repulsion set, introducing spurious separation that pushes apart semantically aligned items and degrades downstream performance.

\paragraph{Effect of HaMR.}
To evaluate the role of collision reduction, we remove HaMR from QuaSID (denoted as \emph{w/o HaMR}). As shown in Table~\ref{tab:ablation}, this variant consistently underperforms the full model across both datasets and evaluation metrics. These results indicate that explicitly penalizing unreasonable SID collisions is critical for preserving semantic discriminability in the discrete token space.

\subsection{Hyperparameter Sensitivity Analysis}

We evaluate QuaSID’s sensitivity to key hyperparameters on Amazon-Beauty, focusing on the collaborative contrastive weight $\lambda_{cl}$ and the repulsion weights for full and partial SID collisions, $\lambda_{\text{full}}$ and $\lambda_{\text{partial}}$. Each experiment varies one hyperparameter while keeping the others at default values. Figure~\ref{fig:hp_sensitive} reports NDCG@K (bar plots, left y-axis) and HR@K (line plots, right y-axis).

\paragraph{Effect of $\lambda_{cl}$.}
$\lambda_{cl}$ controls the strength of collaborative supervision that aligns SIDs with behavioral signals. Increasing $\lambda_{cl}$ from 0.01 to 0.1 consistently improves performance, with HR@10 and NDCG@10 peaking around 0.1, suggesting that moderate contrastive regularization complements reconstruction/quantization by injecting collaborative semantics. Further increasing $\lambda_{cl}$ to 0.5 degrades performance, indicating that overly strong contrastive forces can dominate optimization and conflict with discretization and collision mitigation.

\paragraph{Effect of $\lambda_{\text{full}}$.}
$\lambda_{\text{full}}$ penalizes full SID collisions (zero Hamming distance), i.e., the most severe ambiguity. The best results occur at $\lambda_{\text{full}}=0.2$. Smaller values under-penalize full collisions and reduce ranking quality, while larger values also hurt performance, likely because excessive repulsion distorts the representation space and weakens semantic generalization.

\paragraph{Effect of $\lambda_{\text{partial}}$.}
$\lambda_{\text{partial}}$ weights penalties for near-collisions within a Hamming radius $R=1$. Performance improves as $\lambda_{\text{partial}}$ increases from 0.01 to 0.1, implying that mild near-collision repulsion alleviates local codebook overcrowding. However, larger values (0.5, 0.8) consistently reduce HR and NDCG, suggesting that over-penalizing partial collisions can unnecessarily separate semantically related items that naturally share some tokens.

\section{Conclusion and Future Work}
We investigated SIDs for multimodal items and identified two key obstacles: token collisions and collision-signal heterogeneity during training, where discrete overlaps may conflate harmful conflicts with benign, task-consistent relations. We proposed QuaSID, integrating HaMR for severity-aware margin repulsion, CVPM to filter same-item pairs and constructed collaborative positives from collision supervision, and an auxiliary dual-tower contrastive loss to inject collaboration signals into tokenization. Extensive offline benchmarks and large-scale online A/B tests show consistent gains across metrics. Future work will further study collision-signal heterogeneity by automatically separating benign overlaps from true semantic conflicts and learning task-conditional qualification policies for when (and how strongly) to repel collisions.

\bibliographystyle{ACM-Reference-Format}
\bibliography{sample-base}

\appendix
\section{Additional Experimental Details}

\subsection{Baselines}
\label{app:baselines}
We compare QuaSID with a set of strong vector-quantization-based baselines in our offline tokenizer evaluation. These methods differ in their quantization strategies, codebook optimization mechanisms, and gradient estimation techniques.
\begin{itemize}
    \item \textbf{RQ-VAE}~\cite{DBLP:conf/cvpr/LeeKKCH22} recursively quantizes the residuals of dense item embeddings using multiple stacked codebooks, enabling a coarse-to-fine discretization. It is one of the most commonly adopted frameworks for semantic ID construction.
    \item \textbf{GRVQ} (Grouped Residual Vector Quantization)~\cite{DBLP:journals/corr/abs-2305-02765} partitions the embedding space into multiple groups and applies residual quantization independently within each group, aiming to reduce cross-dimension interference and improve codebook utilization.
    \item \textbf{Improved VQGAN}~\cite{DBLP:conf/iclr/YuLKZPQKXBW22} adopts a low-dimensional codebook and applies $\ell_2$-normalized to both codebook vectors and encoder outputs, effectively using cosine similarity for code assignment. This design encourages more uniform code utilization and improves reconstruction quality.
    \item \textbf{RQ-VAE-Rotation}~\cite{DBLP:conf/iclr/FiftyJDILATR25} replaces the STE in RQ-VAE with the rotation trick, which leads to more informative gradients and improved quantization stability.
    \item \textbf{SimRQ}~\cite{zhu2025addressing} freezes the codebooks and implicitly generates discrete codes via linear projections. This design mitigates codebook collapse and simplifies optimization, while preserving the hierarchical representation benefits of residual quantization.
    \item \textbf{RQ-OPQ}~\cite{DBLP:journals/corr/abs-2509-03236} combines residual quantization with optimized product quantization.
    \item \textbf{RQ-Kmeans}~\cite{DBLP:conf/cikm/LuoCSYHYLZ0HQZZ25} is a two-stage semantic ID learning method that first aligns item representations based on item--item collaborative pairs, and then constructs residual quantization codebooks via a K-means-based mechanism to generate multi-layer discrete codes.
\end{itemize}
All methods use the same backbone model, and differ only in the tokenizer used to construct SIDs.

\subsection{Implementation Details}
\label{app:impl}
We use TIGER~\cite{DBLP:conf/nips/RajputMSKVHHT0S23} as the generative recommendation backbone for all methods and keep identical model configurations across QuaSID and baselines.
All offline experiments are implemented in PyTorch based on the open-source RQ-VAE recommender framework\footnote{https://github.com/EdoardoBotta/RQ-VAE-Recommender}.
We optimize all generative models with Adam (lr $3\times 10^{-4}$, weight decay $1\times 10^{-5}$) and batch size 256.
Early stopping is applied if NDCG@5 + HitRate@5 does not improve for 10 consecutive validation checks.
We adopt an 8-layer Transformer with 8 attention heads, embedding dimension 128, and MLP hidden dimension 512.
All results are averaged over five random seeds.

For offline semantic ID learning, we use batch size 256 for all methods; for online A/B tests, we use batch size 1024.
In offline public-dataset experiments, we set the number of codebook layers to $L=3$ and the codebook size to $K=256$ for a fair comparison across methods.
In industrial experiments, we use $L=4$ and $K=1024$.
Baseline hyperparameters follow the original papers and are further tuned on validation sets.
For QuaSID, we set the Hamming radius to $R=1$ for offline experiments and $R=2$ for online A/B tests, with margins $m_{\text{full}}=0.8$ and $m_{\text{partial}}=0.5$.
We tune $\lambda_{cl}\in[0.01,0.5]$, $\lambda_{\text{full}}\in[0.05,0.8]$, and $\lambda_{\text{partial}}\in[0.01,0.8]$ on validation sets.

\end{document}